\documentclass[twocolumn,secnumarabic,amssymb,nobibnotes,aps,superscriptaddress,pra]{revtex4-1} 

\usepackage[dvips]{graphicx}
\usepackage{bm}
\usepackage{latexsym} 
\usepackage{amsmath}
\usepackage{amssymb}
\usepackage{color}

\newcommand{\ui}{{\rm i}}
\newcommand{\veps}{{\varepsilon}}

\newcommand{\bmH}{{\bm H}}

\newcommand{\bmm}{{\bm m}}
\newcommand{\bmn}{{\bm n}}

\newcommand{\bmS}{{\bm S}}
\newcommand{\bmsig}{{\bm \sigma}}

\newcommand{\bra}{\langle}
\newcommand{\ket}{\rangle}
\newcommand{\kB}{k_{\rm B}}

\newcommand{\bfzhat}{{\bf \hat{z}}}

\makeatletter
\renewcommand*{\p@subsection}{}
\renewcommand*{\p@subsubsection}{}
\makeatother

\begin{document}

\title{
  Spin Seebeck effect in paramagnets and antiferromagnets at elevated temperatures 
}

\author{Yutaka Yamamoto} 
\affiliation{Department of Physics, Okayama University, Okayama 700-8530, Japan}
\author{Masanori Ichioka}
\affiliation{Research Institute for Interdisciplinary Science, Okayama University, Okayama 700-8530, Japan}
\affiliation{Department of Physics, Okayama University, Okayama 700-8530, Japan}
\author{Hiroto Adachi}
\affiliation{Research Institute for Interdisciplinary Science, Okayama University, Okayama 700-8530, Japan}
\affiliation{Department of Physics, Okayama University, Okayama 700-8530, Japan}
\date{\today}

\begin{abstract}
  We develop a theory of the spin Seebeck effect (SSE) in paramagnets as well as in antiferromagnets at elevated temperatures where the classical limit of the fluctuation-dissipation theorem is applicable. Employing dissipative stochastic models that are valid at these temperatures, we calculate the SSE signal, and we find that both the paramagnetic SSE and the antiferromagnetic SSE are expressed by a single equation that is proportional to the external magnetic field times the spin susceptibility of the magnet. The present result suggests the appearance of a cusp structure at the N\'{e}el temperature in the antiferromagnetic SSE signal. 
\end{abstract} 

\pacs{}

\keywords{} 

\maketitle

\section{Introduction \label{Sec:I}}
The spin Seebeck effect (SSE)~\cite{Uchida08,Jaworski10,Uchida10a} refers to the spin injection from a magnet into the adjacent spin-Hall electrode that is driven by a temperature gradient, where no charge transfer across the interface between spin-injecting magnet/spin-Hall electrode is involved~\cite{Bauer12}. While examples of the spin-Hall electrode range from nonmagnetic metals~\cite{Uchida10b,Qu13,Qu14,Vlietstra14,Guo16}, to oxides~\cite{Qiu15,Shiomi15}, to magnetic metals and alloys~\cite{Miao13,Kikkawa13,Mendes14,Seki15,Qu15,Zou16}, a typical choice of the spin-injecting magnet has been one of {\it ferrimagnetic} insulators such as garnet ferrites or spinel ferrites~\cite{Uchida16}. In these ferrimagnets, the magnetization is the order parameter characterizing the magnetic state, and it is customary to consider~\cite{Xiao10,Adachi11,Adachi13} that the main actor for the SSE is the spin wave or the magnon that causes a thermal version of spin pumping. 

Recently, the SSE was measured by choosing a different class of materials other than ferrimagnets as the spin-injecting magnet. In Ref.~\cite{Wu15}, the SSE in paramagnetic insulators Gd$_3$Ga$_5$O$_{12}$ and DyScO$_3$ was reported. More recently, Refs.~\cite{SSeki15,Wu16,Holanda17} reported the SSE in antiferromagnetic insulators Cr$_2$O$_3$, MnF$_2$, and NiO. The crucial difference between these paramagnets and antiferromagnets and the ferrimagnets lies in the following fact. In paramagnets and antiferromagnets, the magnetization is {\it not} the order parameter. In ferrimagnets, by contrast, the magnetization is the broken-symmetry variable, and thus it plays the role of the order parameter. Therefore, from a theoretical point of view, the paramagnetic and antiferromagnetic SSEs have a common feature in that the spin current is injected from a material that does not possess {\it spontaneous} magnetization~\cite{Kittel-intro}. 

In the literature, the antiferromagnetic SSE has been discussed theoretically in several publications~\cite{Ohnuma13,Rezende16,Bender17}. However, these works are justified at low enough temperature well below the N\'{e}el temperature $T_{\rm N}$ ($T \ll T_{\rm N}$). This is because the Holstein-Primakoff boson is used in Refs.~\cite{Ohnuma13,Rezende16}, or the amplitude of the order parameter (staggered magnetization) in the ground state is assumed to be temperature-independent in Ref.~\cite{Bender17}. In this connection, it is worth mentioning that there is a theory dealing with spin transport in paramagnets and antiferromagnets via a Schwinger auxiliary boson/fermion representation~\cite{Okamoto16}, but the theory does not consider the SSE. Therefore, it is of vital importance to develop a theory that is applicable at temperatures both near and above $T_{\rm N}$.

In this paper, we develop a theory of the paramagnetic SSE and antiferromagnetic SSE at elevated temperatures where the classical limit of the fluctuation-dissipation theorem is applicable. For this purpose, we use dissipative stochastic models that have been well established in the field of dynamic critical phenomena~\cite{Chaikin-text}, and were successfully applied to the ferromagnetic SSE near the Curie temperature~\cite{Adachi18}. First, we apply this method to the paramagnetic SSE, where the corresponding dynamic equation is the stochastic Bloch equation~\cite{Adachi13}. In this case, calculation of the paramagnetic SSE can be done in a manner similar to that of the ferromagnetic SSE. Next, we extend the calculation to the antiferromagnetic SSE, where the corresponding dynamic equation is the time-dependent Ginzburg-Landau equation~\cite{Chaikin-text}. Note that calculation of the antiferromagnetic SSE is much more involved than that of the ferromagnetic and paramagnetic SSEs, since in this case the two degrees of freedom (magnetization and staggered magnetization) are tightly coupled by exchange interaction~\cite{Kamra18} such that we need to deal with a complex matrix algebra. Indeed, as one can see in Appendix~\ref{Sec:App01}, a lengthy and tedious calculation is required in order to obtain the result satisfying the second law of thermodynamics, i.e., a signal proportional to the temperature bias.

Starting from these two different models, the paramagnetic SSE and antiferromagnetic SSE are calculated. We find that, despite a marked difference in the model equations, both SSEs are expressed by the same equation, which is proportional to the spin susceptibility of the spin-injecting magnet multiplied by the external magnetic field. From this result, as well as recalling that the spin susceptibility in antiferromagnets has a kink at $T_{\rm N}$, we conclude that a cusp structure appears at $T_{\rm N}$ in the antiferromagnetic SSE signal. 

This paper is organized as follows. In Sec.~\ref{Sec:II}, we develop a theory of the paramagnetic SSE. In Sec.~\ref{Sec:III}, we extend the calculation to the antiferromagnetic SSE near and above $T_{\rm N}$. Finally in Sec.~\ref{Sec:IV}, we discuss and summarize our results.

\section{Paramagnetic spin Seebeck effect \label{Sec:II}}  
In this section, we develop a theory of the paramagnetic SSE, by extending the calculation of the ferromagnetic SSE near the Curie temperature~\cite{Adachi18}. Starting from the stochastic Bloch equation, we calculate the paramagnetic SSE, and we show that the signal is proportional to the spin susceptibility of the spin-injecting magnet multiplied by the external magnetic field. 

\subsection{Model} 
\begin{figure}[t] 
  \begin{center}
    \scalebox{0.4}[0.4]{\includegraphics{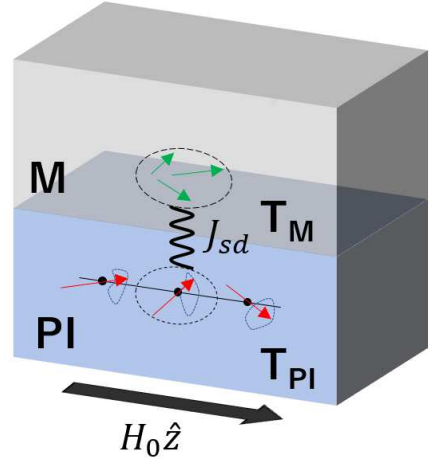}}
  \end{center}
  \caption{Schematic illustration of the system considered in Sec.~\ref{Sec:II} for the paramagnetic SSE. Here, PI and M refer to a paramagnetic insulator and a metal, respectively.}
  \label{fig:model_PISSE}
\end{figure}

We consider a bilayer composed of a paramagnetic insulator (PI) with its temperature $T_{\rm PI}$, and a metal (M) acting as a spin-Hall electrode with its temperature $T_{\rm M}$, as shown in Fig.~\ref{fig:model_PISSE}. Our starting point is the stochastic Bloch equation for localized spin $\bmS$ in PI: 
\begin{equation}
  \frac{\partial}{\partial t} \bmS = \gamma \bmH_0 \times \bmS 
  - \Gamma_{\rm PI} \left( \bmS- \bmS_{\rm eq} \right) 
  + \frac{J_{\rm sd}}{\hbar} \bmsig \times \bmS + {\bm \xi},
  \label{eq:BlochPI}
\end{equation}
where $\gamma$ is the gyromagnetic ratio, $\bmH_0= H_0 \bfzhat$ is the uniform external magnetic field, $\Gamma_{\rm PI}$ is the spin relaxation rate of PI, and $J_{\rm sd}$ is the $s$-$d$ exchange interaction at the PI/M interface. The equilibrium value of $\bmS$ is given by
\begin{equation}
  \bmS_{\rm eq}= \chi_{\rm PI} g \mu_{\rm B} H_0 \bfzhat,
  \label{eq:chiPI01}
\end{equation}
where $g$ is the g-factor, $\mu_{\rm B}$ is the Bohr magneton, and $\chi_{\rm PI}$ is the spin susceptibility of PI. The last term, ${\bm \xi}$, on the right-hand side of Eq.~(\ref{eq:BlochPI}) is the thermal noise field in PI, represented by a Gaussian random variable with zero mean and variance, 
\begin{equation}
  \bra \xi^i(t) \xi^j(t') \ket = 2 \kB T_{\rm PI} \chi_{\rm PI} \Gamma_{\rm PI} \delta_{i,j} \delta (t-t'), 
  \label{eq:xixi} 
\end{equation}
where $\kB$ is the Boltzmann constant, and $\bra \cdots \ket$ means averaging over the thermal noise. 

Similarly, we consider the Bloch equation for the spin density $\bmsig$ in M: 
\begin{equation} 
  \frac{\partial}{\partial t} \bmsig = -\frac{1}{\tau_{\rm M}}
  \left( \bmsig- \chi_{\rm M} J_{\rm sd} \bmS \right) + \frac{J_{\rm sd}}{\hbar} \bmS \times \bmsig + {\bm \zeta},
  \label{eq:BlochSH01}
\end{equation}
where $\tau_{\rm M}$ is the spin relaxation time of M. The equilibrium spin density is given by
\begin{equation}
    \bmsig_{\rm eq}= \chi_{\rm M} J_{\rm sd} \bmS_{\rm eq} 
  \label{eq:bmsig_eq01}
\end{equation}
with $\chi_{\rm M}$ being the spin susceptibility of M. The last term on the right-hand side of Eq.~(\ref{eq:BlochSH01}) is the thermal noise field in M, which is represented by a Gaussian random variable with zero mean and variance 
\begin{equation}
  \bra \zeta^i(t) \zeta^j(t') \ket = \frac{2 \kB T_{\rm M} \chi_{\rm M}}{\tau_{\rm M}}
  \delta_{i,j} \delta (t-t').
  \label{eq:zetazeta}  
\end{equation}

Here are a few comments on our model. First, the spin dephasing of M is assumed to be very strong, so that a precession term of the form $\gamma \bmH_0 \times \bmsig$ is disregarded in Eq.~(\ref{eq:BlochSH01}). Second, consistently with this assumption, the spin relaxation rate of PI is assumed to be much weaker than that of M, i.e., $\Gamma_{\rm PI} \ll \tau_{\rm M}^{-1}$. Third, although there could be a term $\bmS'_{\rm eq}= J_{\rm sd} \chi_{\rm PI} \bmsig_{\rm eq}$ in the equilibrium value of $\bmS$, such a term does not affect the following perturbative calculation with respect to $J_{\rm sd}$. Finally, Eqs.~(\ref{eq:xixi}) and (\ref{eq:zetazeta}) are required by the fluctuation-dissipation theorem, which is derived from the postulate that the equilibrium probability of finding the spin variable equals the Boltzmann distribution~\cite{Chaikin-text}.

\subsection{Spin injection signal}
We calculate the spin current injected into the metal M by the paramagnetic SSE. Following \cite{Adachi18}, we define the spin current $I_s$ as the rate of change of the spin density in M, i.e., $I_s = \frac{\partial}{\partial t}\bra \sigma^z \ket$. Because the SSE is driven by the dynamic fluctuations of $\bmS$ and $\bmsig$~\cite{Adachi13}, it is convenient to introduce their fluctuation components $\delta \bmS = \bmS - \bmS_{\rm eq}$ and $\delta \bmsig = \bmsig - \bmsig_{\rm eq}$. Then, using the $z$-component of the Bloch equation~(\ref{eq:BlochSH01}) and assuming negligibly small spin memory loss at the PI/M interface~\cite{Haney13,Chen15,Zhu19}, $I_s$ is calculated to be 
\begin{equation}
  I_s (t) = \frac{J_{\rm sd}}{\hbar} {\rm Im} \bra \delta S^- (t) \delta \sigma^+ (t) \ket,
  \label{eq:IsBloch01} 
\end{equation}
where the quantity $O^\pm$ of a variable ${\bf O}$ is defined by 
\begin{equation}
  O^\pm = O^x \pm \ui O^y.
  \label{eq:A_pm01}
\end{equation}
We now assume that the system is in the steady state where both sides of Eq.~(\ref{eq:IsBloch01}) are independent of $t$. Then $I_s$ can be represented spectrally as 
\begin{equation}
  I_s = \frac{J_{\rm sd}}{\hbar} \int_\omega {\rm Im} \bra \bra \delta S^-_{\omega} \delta \sigma^+_{-\omega} \ket \ket,
  \label{eq:Ispara01}
\end{equation}
where the Fourier transform of a function $f(t)$ is given by $f(t)= \int_\omega f_\omega e^{-\ui \omega t}$ with the shorthand notation $\int_\omega = \int_{-\infty}^\infty \frac{d \omega}{2 \pi}$. In the above equation, the quantity $\bra \bra \delta S^-_{\omega} \delta \sigma^+_{-\omega} \ket \ket$ is defined by $\bra \delta S^-_{\omega} \delta \sigma^+_{\omega'} \ket = 2 \pi \delta (\omega+ \omega') \bra \bra \delta S^-_{\omega} \delta \sigma^+_{-\omega} \ket \ket$. 

To evaluate the right-hand side of Eq.~(\ref{eq:Ispara01}), we use the perturbative approach with respect to $J_{\rm sd}$, and we expand $\delta S^{\pm}_\omega$ and $\delta \sigma^{\pm}_\omega$ as  
\begin{equation}
  \delta S^\pm = \delta S^{\pm (0)}+ \delta S^{\pm (1)} 
  \label{eq:dS0001}
\end{equation}
and 
\begin{equation}
  \delta \sigma^\pm = \delta \sigma^{\pm (0)}+ \delta \sigma^{\pm (1)}, 
  \label{eq:dsig0001}
\end{equation}
where $\delta S^{\pm (0)}$ and $\delta \sigma^{\pm (0)}$ are independent of $J_{\rm sd}$, whereas $\delta S^{\pm (1)}$ and $\delta \sigma^{\pm (1)}$ are the first-order corrections. Substituting Eqs.~(\ref{eq:dS0001}) and (\ref{eq:dsig0001}) into Eq.~(\ref{eq:Ispara01}) and summarizing the result up to linear order with respect to $J_{\rm sd}$, the injected spin current is written as 
\begin{eqnarray}
  I_s &=& \frac{J_{\rm sd}}{\hbar} \int_\omega
  {\rm Im} \Big[ \bra \bra \delta S^{-(0)}_{\omega} \delta \sigma^{+(1)}_{-\omega} \ket \ket
    + \bra \bra \delta S^{-(1)}_{\omega} \delta \sigma^{+(0)}_{-\omega} \ket \ket \Big]. \quad \;  
  \label{eq:Ispara02} 
\end{eqnarray}
Therefore, the remaining task is to calculate $\delta S^{-(0)}_\omega$, $\delta S^{-(1)}_\omega$, $\delta \sigma^{+(0)}_{-\omega}$, and $\delta \sigma^{+(1)}_{-\omega}$ in order to evaluate Eq.~(\ref{eq:Ispara02}). 

We write the Bloch equation (\ref{eq:BlochPI}) for $\delta S^{\pm}_\omega$: 
\begin{equation}
  \left( \omega \pm \gamma H_0 + \ui \Gamma_{\rm PI} \right) \delta S^\pm_\omega 
  = \pm \frac{J_{\rm sd} S_{\rm eq} }{\hbar} \delta \sigma_\omega^\pm+ \ui \xi^\pm_\omega,
  \label{eq:BlochPI02}
\end{equation}  
and the Bloch equation (\ref{eq:BlochSH01}) for $\delta \sigma^{\pm}_\omega$: 
\begin{equation}
  \left( \omega + \ui \tau_{\rm M}^{-1} \right) \delta \sigma^\pm_\omega
  = \ui \frac{\chi_{\rm M}J_{\rm sd}}{\tau_{\rm M}} \delta S^{\pm}_\omega+ \ui \zeta^\pm_\omega.
    \label{eq:BlochSH02}
\end{equation}
From Eqs.~(\ref{eq:BlochPI02}) and (\ref{eq:BlochSH02}), the unperturbed solutions are obtained as 
\begin{equation}
    \delta S^{-(0)}_{\omega} = G(\omega) \ui \xi^-_{\omega}, 
  \label{eq:dS00}
\end{equation}
and 
\begin{equation}
  \delta \sigma^{+(0)}_{-\omega} = -g^*(\omega) \ui \zeta^+_{-\omega},
  \label{eq:dsig00}
\end{equation}
where $G(\omega)= (\omega- \gamma H_0+ \ui \Gamma_{\rm PI})^{-1}$ and $g(\omega)= (\omega + \ui \tau^{-1}_{\rm M})^{-1}$. In a similar way, the first-order corrections $\delta S^{-(1)}$ and $\delta \sigma^{+(1)}$ are given by 
\begin{equation}
  \delta S^{-(1)}_{\omega} = -\frac{J_{\rm sd}}{\hbar}S_{\rm eq} G(\omega) g(\omega) \ui \zeta^+_{\omega}, 
  \label{eq:dS01}
\end{equation}
and
\begin{equation}
  \delta \sigma^{+(1)}_{-\omega} =
  \ui \frac{J_{\rm sd} \chi_{\rm M}}{\tau_{\rm M}}
  g^*(\omega) G^*(\omega) \ui \xi^+_{-\omega}. 
  \label{eq:dsig01}
\end{equation}

Substituting Eqs.~(\ref{eq:dS00})--(\ref{eq:dsig01}) into Eq.~(\ref{eq:Ispara02}), the injected spin current is calculated to be
\begin{equation}
  I_s = I_s^{\rm pump}- I_s^{\rm back}, 
  \label{eq:Ispara03}
\end{equation}
where
\begin{equation}
  I_s^{\rm pump} = -\frac{J_{\rm sd}^2}{\hbar} \int_\omega
  |G(\omega)|^2 |g(\omega)|^2 \omega \frac{\chi_{\rm M}}{\tau_{\rm M}} 
  \bra \bra \xi^{-}_{\omega} \xi^{+}_{-\omega} \ket \ket 
  \label{eq:Ipump01}
\end{equation}
and 
\begin{equation}
  I_s^{\rm back}= -\frac{J_{\rm sd}^2}{\hbar} \int_\omega
    |G(\omega)|^2 |g(\omega)|^2 \frac{S_{\rm eq}\Gamma_{\rm PI}}{\hbar} 
  \bra \bra \zeta^{-}_{\omega} \zeta^{+}_{-\omega} \ket \ket.
  \label{eq:Iback01}
\end{equation}

To proceed further, we first use the spectral representations of Eqs.~(\ref{eq:xixi}) and (\ref{eq:zetazeta}), which reduce to $\bra \bra \xi^{-}_{\omega} \xi^{+}_{-\omega} \ket \ket = 4 \kB T_{\rm PI} \chi_{\rm PI} \Gamma_{\rm PI}$ and $\bra \bra \zeta^{-}_{\omega} \zeta^{+}_{-\omega} \ket \ket = 4 \kB T_{\rm M} \chi_{\rm M} /\tau_{\rm M}$ in the present case. Next, we evaluate the integral over $\omega$ by picking up the magnon pole $\omega= \gamma H_0+ \ui \Gamma_{\rm PI}$, which is justified by the assumption $\Gamma_{\rm PI} \ll \tau_{\rm M}^{-1}$ mentioned below Eq.~(\ref{eq:zetazeta}). After evaluating the residue at the magnon pole, we finally obtain 
\begin{equation}
  I_s =
  - \frac{2 J_{\rm sd}^2 \tau_{\rm M} \chi_{\rm M}}{\hbar^2 } 
  S_{\rm eq} \kB (T_{\rm PI}- T_{\rm M}) .
  \label{eq:Ispara_fin01}
\end{equation}
Using the relation $S_{\rm eq}/\hbar= \chi_{\rm PI} \gamma H_0$, the above result can be rewritten as 
\begin{equation}
  I_s =
  - \frac{2 J_{\rm sd}^2 \tau_{\rm M} \chi_{\rm M}}{\hbar } 
  \chi_{\rm PI} \gamma H_0 \kB \Delta T, 
  \label{eq:Ispara_fin02}
\end{equation}
where we introduced the notation $\Delta T= T_{\rm PI}- T_{\rm M}$. 

Equation~(\ref{eq:Ispara_fin02}) shows that the paramagnetic SSE is proportional to the spin susceptibility $\chi_{\rm PI}$ of PI, multiplied by the external magnetic field $H_0$. This means that the calculated paramagnetic SSE signal is proportional to the field-induced magnetization in PI. Note that this result is consistent with the experimental result reported in Ref.~\cite{Wu15}. Later, Eq.~(\ref{eq:Ispara_fin02}) is used to argue that the paramagnetic SSE and the antiferromagnetic SSE are expressed by a single equation. 

\section{Antiferromagnetic spin Seebeck effect \label{Sec:III}}
In this section, we develop a theory of the antiferromagnetic SSE near and above $T_{\rm N}$. Starting from the time-dependent Ginzburg-Landau equation for a uniaxial antiferromagnet, we calculate the antiferromagnetic SSE and show that the signal is proportional to the spin susceptibility of the antiferromagnet, multiplied by the external magnetic field. As noted in the Introduction, the calculation is much more involved than that of the previous section, since in this case the two degrees of freedom (magnetization and staggered magnetization) are tightly coupled by exchange interaction~\cite{Kamra18}, such that a complex matrix algebra is required.  

\begin{figure}[t] 
  \begin{center}
    \scalebox{0.4}[0.4]{\includegraphics{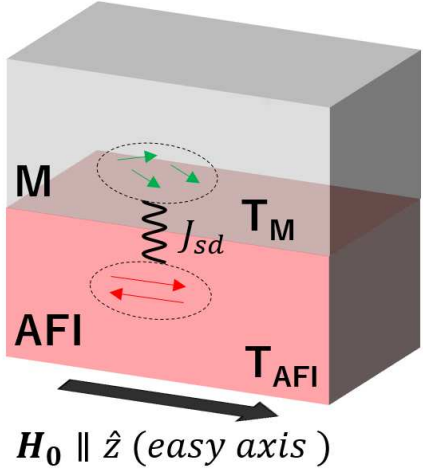}}
  \end{center}
  \caption{Schematic illustration of the system considered in Sec.~\ref{Sec:III} for the antiferromagnetic SSE. Here, AFI and M refer to an antiferromagnetic insulator and a metal, respectively.}
  \label{fig:model_AFISSE}
\end{figure}

\subsection{Model}
We consider a bilayer composed of an antiferromagnetic insulator (AFI) with its temperature $T_{\rm AFI}$ and a metal (M) with its temperature $T_{\rm M}$, as shown in Fig.~\ref{fig:model_AFISSE}. For AFI, we use the Ginzburg-Landau free energy of the following form~\cite{Landau-elec}: 
\begin{eqnarray}
  F_{\rm GL} &=& \veps_0 \int d^3 r \Bigg\{ \frac{u}{2} \bmn^2+ \frac{v}{4} \left( \bmn^2 \right)^2
  + \frac{K}{2} \left( \bmn \times \bfzhat \right)^2   \qquad \qquad \nonumber \\
  &+& \frac{r_0}{2} \bmm^2 + \frac{w}{2} \bmm^2 \bmn^2 
   - \frac{{\bmH}_0}{\mathfrak{h}_0} \cdot \bmm \Bigg\}
   - J_{\rm sd} \bmsig \cdot \bmm, 
\end{eqnarray}
where $\bmm$ and $\bmn$ are respectively the total and staggered spins which are coarse-grained within an effective cell volume $v_0$, and $\veps_0 = \mathfrak{h}_0^2$ is the magnetic energy density with $\mathfrak{h}_0= \gamma \hbar/v_0$. In the above equation, the first three terms on the right hand side describe the physics of staggered spin $\bmn$, where $u= (T-T_{\rm N})/T_{\rm N}$ measures the distance from the N\'{e}el Temperature, $v$ is the quartic term coefficient, $K$ is the uniaxial anisotropy constant, and the gradient term is disregarded because the spatial fluctuation does not change the main result as in the case of the ferromagnetic SSE near the Curie temperature~\cite{Adachi18}. The fourth term concerns the total spin $\bmm$, where $r_0^{-1}= A/(T+ \Theta)$ with two parameters $A$ and $\Theta$ is the paramagnetic spin susceptibility of AFI at $T > T_{\rm N}$ in the dimensionless form. The fifth term comes from the interaction between $\bmm$ and $\bmn$~\cite{Landau-elec}, and the sixth term is the coupling between $\bmm$ and a static external magnetic field $\bmH_0= {H}_0 \bfzhat$ applied paralell to the easy axis. The last term represents the coupling of $\bmm$ to the spin density $\bmsig$ through the $s$-$d$ interaction $J_{\rm sd}$ at the AFI/M interface. Note that the strength of the external magnetic field is assumed to be much smaller than the spin-flop critical field, so that the spin-flop transition is not considered here.

Following \cite{Freedman76} and \cite{Halperin76}, we consider the time-dependent Ginzburg-Landau dynamics for AFI:
\begin{eqnarray}
  \frac{\partial}{\partial t} \bmm &=& \gamma \bmH_m \times \bmm + \gamma \bmH_n \times \bmn
  + \Gamma_m \bmH_m + {\bm \xi}, \label{eq:TDGL_m01} \\
  \frac{\partial}{\partial t} \bmn &=& \gamma \bmH_n \times \bmm + \gamma \bmH_m \times \bmn 
  + \Gamma_n \bmH_n + {\bm \eta}, \label{eq:TDGL_n01}
\end{eqnarray}
where $\Gamma_m $ and $\Gamma_n $ are dissipation coefficients. The effective fields $\bmH_m$ and $\bmH_n$ are defined by 
\begin{equation}
  \bmH_m = - \frac{1}{\mathfrak{h}_0} \frac{\delta F_{\rm GL}}{\delta \bmm}
\end{equation}
and
\begin{equation}
  \bmH_n = - \frac{1}{\mathfrak{h}_0} \frac{\delta F_{\rm GL}}{\delta \bmn}. 
\end{equation}
In Eqs.~(\ref{eq:TDGL_m01}) and (\ref{eq:TDGL_n01}), the two fields ${\bm \xi}$ and ${\bm \eta}$ represent thermal noises for $\bmm$ and $\bmn$, taking the form of Gaussian white noises with zero means and variances: 
\begin{equation}
  \bra \xi^i(t) \xi^j(t') \ket = \frac{2 \kB T_{\rm AFI} \Gamma_m}{\veps_0 v_0} \delta_{i,j} \delta (t-t') 
  \label{eq:ximxim}
\end{equation}
and 
\begin{equation}
  \bra \eta^i(t) \eta^j(t') \ket = \frac{2 \kB T_{\rm AFI} \Gamma_n}{\veps_0 v_0} \delta_{i,j} \delta (t-t').  
  \label{eq:etaeta}
\end{equation}
Note that the two noises ${\bm \xi}$ and ${\bm \eta}$ are assumed to be statistically independent, such that they satisfy 
\begin{equation}
  \bra \xi^i(t) \eta^j(t') \ket = 0. 
  \label{eq:ximeta} 
\end{equation}

\begin{figure}[t] 
  \begin{center}
    \scalebox{0.5}[0.5]{\includegraphics{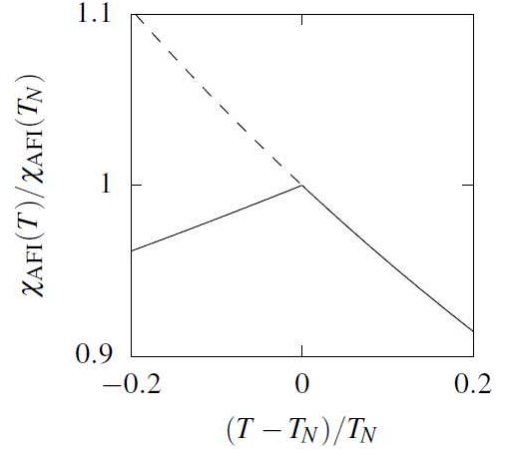}}
  \end{center}
  \caption{Static spin susceptibility [Eq.~(\ref{eq:chiAFI01})] calculated for AFI as a function of temperature $T$. Here, $A/T_{\rm N}= 0.143$, $\Theta/T_{\rm N}= 1.14$, $v=1.0$, and $w= 10.0$ are used. The dashed line is an extrapolation to lower temperatures assuming no antiferromagnetic order. 
  }
  \label{fig:chi_AFI}
\end{figure}

First, we consider thermal equilibrium of AFI in the absence of $J_{\rm sd}$. The equilibrium value of $\bmn$ is determined by the condition $\bmH_n= {\bm 0}$, which yields $\bmn_{\rm eq}= n_{\rm eq} \bfzhat$ with 
\begin{equation}
  n_{\rm eq} =
  \begin{cases}
    \sqrt{\frac{|u|}{v}} = \sqrt{\frac{T_{\rm N}-T}{T_{\rm N}} \frac{1}{v}} & (T<T_{\rm N}) \\
    0 & (T>T_{\rm N}) 
  \end{cases}   
\end{equation}
due to the uniaxial anisotropy. In deriving the above result, we assumed that the equilibrium value of $\bmm$ is much smaller than that of $\bmn$, i.e., $m_{\rm eq} \ll n_{\rm eq}$, such that a small correction to $n_{\rm eq}$, which is proportional to $w m_{\rm eq}^2$, can be neglected~\cite{Landau-elec}. In line with this assumption, the equilibrium value of the total spin $\bmm_{\rm eq}= m_{\rm eq} \bfzhat$, which is determined by the condition $\bmH_m= {\bm 0}$, is given by 
\begin{equation}
  m_{\rm eq} = \frac{1}{r} \widetilde{H}_0,
  \label{eq:m_eq01}
\end{equation}
where we introduced the normalized magnetic field $\widetilde{H}_0= H_0/\mathfrak{h}_0$, and 
\begin{equation}
  r = r_0+ w n_{\rm eq}^2. 
\end{equation}
Note that Eq.~(\ref{eq:m_eq01}) can be rewritten in the same form as Eq.~(\ref{eq:chiPI01}): 
\begin{equation}
  m_{\rm eq} = \chi_{\rm AFI} \; g\mu_{\rm B} H_0,
  \label{eq:m_eq02}
\end{equation}
where we defined the spin susceptibility of AFI by 
\begin{equation}
  \chi_{\rm AFI} = \frac{1}{r \veps_0 v_0}.
  \label{eq:chiAFI01}
\end{equation}
In Fig.~\ref{fig:chi_AFI}, we plot the calculated $\chi_{\rm AFI}$ as a function of temperature $T$ near $T_{\rm N}$, where the development of the staggered spin reduces the susceptibility~\cite{Landau-elec}. 

As for M, the physics is described by the spin density $\bmsig$, which obeys the Bloch equation of the same form as (\ref{eq:BlochSH01}): 
\begin{equation} 
  \frac{\partial}{\partial t} \bmsig = -\frac{1}{\tau_{\rm M}} \left( \bmsig- \chi_{\rm M} J_{\rm sd} \bmm \right) + \frac{J_{\rm sd}}{\hbar} \bmm \times \bmsig + {\bm \zeta}, 
  \label{eq:BlochSH03}
\end{equation}
where the thermal noise field ${\bm \zeta}$ obeys the same Gaussian ensemble as Eq.~(\ref{eq:zetazeta}). Besides, the equilibrium spin density is given by 
\begin{equation}
    \bmsig_{\rm eq}= \chi_{\rm M} J_{\rm sd} \bmm_{\rm eq}, 
  \label{eq:bmsig_eq02}
\end{equation}
which is essentially the same as Eq.~(\ref{eq:bmsig_eq01}).

Next, we consider nonequilibrium fluctuations of $\bmm$, $\bmn$, and $\bmsig$ by introducing the following decompositions:
\begin{eqnarray}
  \bmm &=& \bmm_{\rm eq} + \delta \bmm, \\
  \bmn &=& \bmn_{\rm eq} + \delta \bmn, \\
  \bmsig &=& \bmsig_{\rm eq} + \delta \bmsig. 
\end{eqnarray}
Then, going into the frequency space as well as using the representation of Eq.~(\ref{eq:A_pm01}), the time-dependent Ginzburg-Landau equation of $\delta \bmm$ and $\delta \bmn$ for the minus branch is summarized as 
\begin{eqnarray}
  \left( \omega - \widehat{\cal A} \right) \begin{pmatrix}
    \delta m^-_{\omega} \\
    \delta n^-_{\omega} 
  \end{pmatrix}
  &=&
  -
  \frac{J_{\rm sd} m_{\rm eq}}{\hbar}
  \begin{pmatrix}
    \delta \sigma^-_{\omega}\\
    0 
  \end{pmatrix} 
  + 
  \begin{pmatrix}
    \ui \xi^-_{\omega}\\    
    \ui \eta^-_{\omega}
  \end{pmatrix},   
  \label{eq:TDGL_minus01}
\end{eqnarray}
where each component of the matrix 
\begin{equation}
  \widehat{\cal A} = \begin{pmatrix}
    a, & b\\
    c, & d
  \end{pmatrix}
\end{equation}
is given by
\begin{eqnarray}
  a &=& \gamma H_0- \ui \Gamma_m r, \\
  b &=& \gamma \mathfrak{h}_0 K n_{\rm eq}, \\
  c &=& \gamma \mathfrak{h}_0 r \, n_{\rm eq}, \\
  d &=& \gamma \mathfrak{h}_0 K m_{\rm eq}- \ui \Gamma_n K. 
\end{eqnarray}
Similarly, the time-dependent Ginzburg-Landau equation of $\delta \bmm$ and $\delta \bmn$ for the plus branch is written as
\begin{eqnarray}
  \left( \omega + \widehat{\cal A}^* \right) \begin{pmatrix}
    \delta m^+_{\omega} \\
    \delta n^+_{\omega} 
  \end{pmatrix}
  &=&
  \frac{J_{\rm sd} m_{\rm eq}}{\hbar}
  \begin{pmatrix}
    \delta \sigma^+_{\omega}\\
    0 
  \end{pmatrix} 
  + 
  \begin{pmatrix}
    \ui \xi^+_{\omega}\\    
    \ui \eta^+_{\omega}
  \end{pmatrix}. 
  \label{eq:TDGL_plus01}
\end{eqnarray}
Finally, the Bloch equation for $\delta \bmsig$ is written as
\begin{equation}
  \left( \omega + \ui \tau_{\rm M}^{-1} \right) \delta \sigma^\pm_\omega =  
  \ui \frac{\chi_{\rm M}J_{\rm sd}}{\tau_{\rm M}} \delta m^\pm_\omega
  + \ui \zeta^\pm_{\omega}.
\end{equation}

\begin{figure}[t] 
  \begin{center}
    \scalebox{1.2}[1.2]{\includegraphics{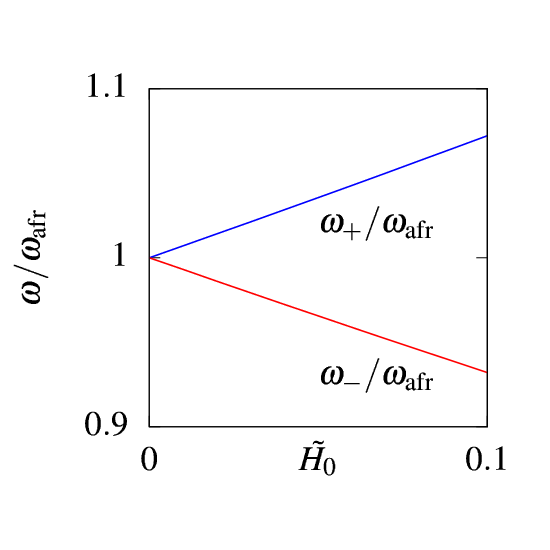}}
  \end{center}
  \caption{Antiferromagnetic resonance frequency $\omega_\pm$ [Eq.~(\ref{eq:omegaPM01})] calculated for AFI as a function of the external magnetic field $\widetilde{H}_0= H_0/\mathfrak{h}_0$, where the frequency is renormalized by $\omega_{\rm afr}= \gamma \mathfrak{h}_0 n_{\rm eq} \sqrt{Kr}$. Here, $A/T_{\rm N}= 0.143$, $\Theta/T_{\rm N}=1.14$, $T/T_{\rm N}= 0.928$, $v=1.0$, $w=10.0$, and $K= 0.5$ are used.}
  \label{fig:AFresonance}
\end{figure}

Let us first discuss the spectrum of spin waves in the present model. For this purpose, we consider Eq.~(\ref{eq:TDGL_minus01}) and set $J_{\rm sd}=0$. Then, the dynamics of $\delta m^-$ and $\delta n^-$ is described by the propagator 
\begin{eqnarray}
  \widehat{\cal G} &=& \left( \omega-\widehat{\cal A} \right)^{-1} \\
  &=&
  \frac{1}{(\omega- \lambda_+)(\omega- \lambda_-)}
  \begin{pmatrix}
    \omega-d, & b \\
    c, & \omega- a 
  \end{pmatrix}, 
  \label{eq:omegaA01}
\end{eqnarray}
where 
\begin{equation}
  \lambda_\pm = \frac{a+ d \pm \sqrt{(a-d)^2+ 4bc} }{2} 
\end{equation}
are the eigenvalues of the two spin wave modes. Now we define
\begin{equation}
  \omega_\pm = \pm {\rm Re}\lambda_\pm.
  \label{eq:omegaPM01}
\end{equation}
Then, in the limit of $H_0=0$, $\omega_\pm$ are given by 
\begin{equation}
  \omega_\pm (H_0 = 0) = \omega_{\rm afr}, 
  \label{eq:Kittel01}
\end{equation}
where we defined $\omega_{\rm afr}= \gamma \mathfrak{h}_0 n_{\rm eq} \sqrt{Kr}$. Equation~(\ref{eq:Kittel01}) coincides with the well-known antiferromagnetic resonance frequency~\cite{Nagamiya51,Kittel51} represented within the Ginzburg-Landau framework (see Eq.~(74.12) in \cite{Landau-stat}). In Fig.~\ref{fig:AFresonance}, we plot $\omega_\pm$ as a function of the external magnetic field $H_0$.

\subsection{Spin injection signal}

The spin current $I_s = \bra \frac{\partial}{\partial t} \sigma^z(t) \ket$ injected into the metal M can be obtained by the $z$-component of the Bloch equation (\ref{eq:BlochSH03}): 
\begin{equation}
  I_s = \frac{J_{\rm sd}}{\hbar} \int_\omega {\rm Im} \bra \bra \delta m^-_{\omega} \delta \sigma^+_{-\omega} \ket \ket, 
  \label{eq:IsAFI01}
\end{equation}
where the steady-state solution is assumed. As in the previous section, we expand $\delta m^\pm$, $\delta n^\pm$, and $\delta \sigma^\pm$ in powers of $J_{\rm sd}$ as
\begin{eqnarray}
  \delta m^\pm &=& \delta m^{\pm (0)}+ \delta m^{\pm (1)}, \\
  \delta n^\pm &=& \delta n^{\pm (0)}+ \delta n^{\pm (1)}, \\
  \delta \sigma^\pm &=& \delta \sigma^{\pm (0)}+ \delta \sigma^{\pm (1)}, 
\end{eqnarray}
where $\delta m^{\pm (0)}$, $\delta n^{\pm (0)}$, and $\delta \sigma^{\pm (0)}$ are independent of $J_{\rm sd}$, whereas $\delta m^{\pm (1)}$, $\delta n^{\pm (1)}$, and $\delta \sigma^{\pm (1)}$ are the first order corrections with respect to $J_{\rm sd}$. Then, up to the linear order with respect to $J_{\rm sd}$, Eq.~(\ref{eq:IsAFI01}) becomes 
\begin{eqnarray}
  I_s &=& \frac{J_{\rm sd}}{\hbar} \int_\omega
  {\rm Im}  \Big[ \bra \bra \delta m^{-(0)}_{\omega} \delta \sigma^{+(1)}_{-\omega} \ket \ket
      + \bra \bra \delta m^{-(1)}_{\omega} \delta \sigma^{+(0)}_{-\omega} \ket \ket \Big] . \; \; \; \; 
    \label{eq:IsAFI02} 
\end{eqnarray}
To proceed further, we need to calculate $\delta m^{-(0)}_{\omega}$, $\delta m^{-(1)}_{\omega}$, $\delta \sigma^{+(0)}_{-\omega}$, and $\delta \sigma^{-(1)}_{-\omega}$. 

The fluctuation $\delta m^-$ of the total spin can be obtained by operating the propagator $\widehat{\cal G}$ to Eq.~(\ref{eq:TDGL_minus01}) from the left. Then, the unperturbed solution is given by 
\begin{equation}
  \delta m^{-(0)}_{\omega}
  =
  G_m (\omega) \ui \xi^-_\omega +   G_n (\omega) \ui \eta^-_\omega,
  \label{eq:dm0AFI01} 
\end{equation}
where
\begin{eqnarray}
  G_m(\omega) &=& \frac{\omega-d}{(\omega- \lambda_+)(\omega- \lambda_-)}, \label{eq:Gmdef01}\\
  G_n(\omega) &=& \frac{b}{(\omega- \lambda_+)(\omega- \lambda_-)} \label{eq:Gndef01}.
\end{eqnarray}
The unperturbed solution for $\delta \sigma^\pm$ is exactly the same as in the previous section, which is given by 
\begin{equation}
  \delta \sigma^{\pm(0)}_{\omega} = g(\omega) \ui \zeta^\pm_\omega,
  \label{eq:dsig0AFI01}
\end{equation}
where $g(\omega)$ is defined below Eq.~(\ref{eq:dsig00}). Similarly, the first-order corrections can be calculated to be
\begin{eqnarray}
  \delta m^{-(1)}_\omega &=& -\frac{J_{\rm sd} m_{\rm eq}}{\hbar} g(\omega) G_m(\omega)
  \ui \zeta^-_{\omega},   \label{eq:dm1AFI01} \\
  \delta \sigma^{+(1)}_{-\omega} &=& \ui \frac{\chi_{\rm M} J_{\rm sd}}{\tau_{\rm M}}
  g^*(\omega) \Big[ G^*_m(\omega) \ui \xi^+_{-\omega} + G^*_n(\omega) \ui \eta_{-\omega}^+ \Big]. \quad 
  \label{eq:dsig1AFI01}
\end{eqnarray}

Substituting Eqs.~(\ref{eq:dm0AFI01}), (\ref{eq:dsig0AFI01}), (\ref{eq:dm1AFI01}), and (\ref{eq:dsig1AFI01}) into Eq.~(\ref{eq:IsAFI02}), the spin current $I_s$ injected into M is calculated to be
\begin{equation}
  I_s = I_{s}^{\rm pump} -I_{s}^{\rm back}, 
\end{equation}
where the pumping current is given by 
\begin{eqnarray}
  I_{s}^{\rm pump} &=& -\frac{J_{\rm sd}^2 \chi_{\rm M}}{\hbar \tau_{\rm M}} \int_\omega
  |g(\omega)|^2 \omega \Big\{ |G_m(\omega)|^2  
  \bra \bra \xi^{-}_{\omega} \xi^{+}_{-\omega} \ket \ket,  \nonumber \\
  && \qquad \qquad \qquad \qquad +  |G_n(\omega)|^2 
  \bra \bra \eta^{-}_{\omega} \eta^{+}_{-\omega} \ket \ket \Big\},   \label{eq:IpumpAFI01}
\end{eqnarray}
whereas the backflow current is 
\begin{eqnarray}
  I_{s}^{\rm back} &=& \frac{J_{\rm sd}^2 m_{\rm eq}}{\hbar^2} \int_\omega 
  |g(\omega)|^2 {\rm Im}G_m(\omega) 
  \bra \bra \zeta^{-}_{\omega} \zeta^{+}_{-\omega} \ket \ket. 
  \label{eq:IbackAFI01}
\end{eqnarray}
Note that the pumping current in Eq.~(\ref{eq:IpumpAFI01}) consists of two terms proportional to $ |G_m(\omega)|^2$ and $ |G_n(\omega)|^2$. This appears consistent with the result of Ref.~\cite{Cheng14}, where it is argued that the spin pumping in antiferromagnets contains two terms proportional to $\bmm \times \dot{\bmm}$ and $\bmn \times \dot{\bmn}$. 

The remaining integral over $\omega$ requires a quite long algebra with the details summarized in Appendix~\ref{Sec:App01}, but the final result is very simple. Following each step explained in Appendix~\ref{Sec:App01} and after a lengthy calculation, the pumped and the backflow currents are respectively calculated to be
\begin{eqnarray}
  I_{s}^{\rm pump} &=& - \frac{2J_{\rm sd}^2 \chi_{\rm M} \tau_{\rm M}}{\hbar^2} m_{\rm eq} \kB T_{\rm AFI}, \label{eq:IpumpAFI02}\\
  I_{s}^{\rm back} &=&  - \frac{2J_{\rm sd}^2 \chi_{\rm M} \tau_{\rm M}}{\hbar^2} m_{\rm eq} \kB T_{\rm M}.  \label{eq:IbackAFI02}
\end{eqnarray}
Using the relation in Eq.~(\ref{eq:m_eq02}) and introducing the notation $\Delta T= T_{\rm AFI}- T_{\rm M}$, the above result can be summarized as 
\begin{equation}
  I_{s} = - \frac{2 J_{\rm sd}^2 \chi_{\rm M} \tau_{\rm M}}{\hbar} \chi_{\rm AFI} \gamma H_0 \kB \Delta T,
  \label{eq:IsAFI_fin01}
\end{equation}
where the relation $\gamma \hbar = g \mu_{\rm B}$ is used.

Equation~(\ref{eq:IsAFI_fin01}) means that the antiferromagnetic SSE is proportional to the external magnetic field $H_0$ times the spin susceptibility $\chi_{\rm AFI}$ of AFI, the form of which is exactly the same as that of the paramagnetic SSE [Eq.~(\ref{eq:Ispara_fin02})].

\section{Discussion and Conclusion \label{Sec:IV}}
The main result of the present paper is that the expressions of the spin injection signal for the paramagnetic SSE [Eq.~(\ref{eq:Ispara_fin02})] and the antiferromagnetic SSE [Eq.~(\ref{eq:IsAFI_fin01})] are the same, and both are proportional to the external magnetic field, multiplied by the spin susceptibility of the magnets. The former result, i.e., the signal being proportional to the external magnetic field in both SSEs at low fields, is consistent with two experiments reported by Wu {\it et al.}~\cite{Wu15,Wu16}. Obviously, the signal vanishes in the absence of the external magnetic field. Turning to the latter result that the signal is proportional to the spin susceptibility of the magnet, the temperature dependence of the spin susceptibility of paramagnets [$\chi_{\rm PI}(T)$] is rather featureless, whereas that of antiferromagnets [$\chi_{\rm AFI}(T)$] shows a kink at $T_{\rm N}$ as the staggered spin reduces the susceptibility below $T_{\rm N}$ (see Fig.~\ref{fig:chi_AFI}). Therefore, the present result indicates that a cusp structure appears at $T_{\rm N}$ in the antiferromagnetic SSE signal.

The importance of the above theoretical result and its impact on future experiments can be summarized as follows. First, the present work enabled a detailed comparison between the theory and experiments for the paramagnetic/antiferromagnetic SSE at elevated temperatures. Second, the calculation revealed that the spin susceptibility is intimately related to the paramagnetic/antiferromagnetic SSE. This means that we can measure the spin susceptibility of a thin film paramagnet/antiferromagnet through the SSE. Thus, we predict that the larger the spin susceptibility of the magnet, the larger is the paramagnetic/antiferromagnetic SSE.

Let us comment on the low-temperature enhancement of the antiferromagnetic SSE observed in \cite{Wu16}. First, although the present theory can properly describe the antiferromagnetic SSE near and above $T_{\rm N}$, calculation of $I_s (T)$ over a wide range of temperatures especially at low temperatures is beyond our scope. This is because the theory is based on the Ginzburg-Landau approach, which is valid only near $T_{\rm N}$, and it uses a high-temperature (classical) limit of the fluctuation-dissipation theorem, which is justified at temperatures above the energy gap of the antiferromagnetic magnons ($\sim$ a few tens kelvin in MnF$_2$). Second, the possible origin of the low-temperature peak has been discussed in Ref.~\cite{Rezende16} in terms of the cancellation of two high-energy magnons with different helicities, or in Ref.~\cite{Adachi10} in terms of phonon drag. A precise measurement of the antiferromagnetic SSE using the technique reported in Ref.~\cite{Iguchi17} would be able to distinguish the true origin of the low-temperature enhancement. 

Let us also remark on the sign of the SSE signal. The present theory concludes the same sign for both the paramagnetic and antiferromagnetic SSEs. This sign is also equal to the SSE in a simple ferromagnet~\cite{Adachi18}.

Before ending, we briefly discuss the antiferromagnetic SSE with an ``uncompensated'' interface. So far, we assumed that the AFI/M interface is atomically rough and the magnetic moment is compensated there, such that there is no net magnetization at the interface. In this situation, Eq.~(\ref{eq:IsAFI_fin01}) applies to both the $a$-type antiferromagnet and the $g$-type antiferromagnet~\cite{Daniels15}. On the other hand, idealistically, we can think of an atomically sharp uncompensated AFI/M interface where a nonzero net magnetization remains at the interface, which may be realized in the $a$-type antiferromagnet. In such a situation, there appears a coupling between $\bmsig$ and $\bmn$ at the interface, and we would expect the appearance of a nonzero antiferromagnetic SSE signal even in the absence of the external magnetic field. 

To summarize, on the basis of the dissipative stochastic models, we have developed a theory of the SSE in paramagnets and antiferromagnets at elevated temperatures. For the paramagnetic SSE, we use the stochastic Bloch equation. For the antiferromagnetic SSE, by contrast, we use the time-dependent Ginzburg-Landau equation. Starting from these two different models we have found that, despite a marked difference in the model equations, both the paramagnetic SSE and the antiferromagnetic SSE are expressed by a single equation, which is proportional to the external magnetic field times the spin susceptibility of the spin-injecting magnet. Moreover, we have clarified that a cusp structure appears at $T_{\rm N}$ in the antiferromagnetic SSE, because of the fact that the antiferromagnetic spin susceptibility has a kink at this point. We hope that our theoretical result is tested experimentally by a careful measurement for the antiferromagnetic SSE.

After completing this work, we became aware of a recent paper, in which the antiferromagnetic SSE in epitaxial FeF$_2$ films is measured~\cite{Li2019}. The experimental data support our theoretical result, since a clear cusp structure at $T_{\rm N}$ is observed in the SSE signal.

\acknowledgments 
We are grateful to Kaya Kobayashi for usuful comment and discussion, and Satoshi Okamoto for comment on the summary of Ref.~\cite{Okamoto16}. This work was financially supported by JSPS KAKENHI Grant No. 15K05151. 

\appendix

\section{Derivation of Eqs.~(\ref{eq:IpumpAFI02}) and (\ref{eq:IbackAFI02})} \label{Sec:App01}
In this appendix, we present the derivation of Eq.~(\ref{eq:IpumpAFI02}) for $I_s^{\rm pump}$ and Eq.~(\ref{eq:IbackAFI02}) for $I_s^{\rm back}$. Because the calculation of the latter is easier, we first deal with Eq.~(\ref{eq:IbackAFI02}). 

We begin with $I_s^{\rm back}$ in Eq.~(\ref{eq:IbackAFI01}), and recall the noise correlator in Eq.~(\ref{eq:zetazeta}) with its spectral representation given by $\bra \bra \zeta^{-}_{\omega} \zeta^{+}_{-\omega} \ket \ket = 4 \kB T_{\rm M} \chi_{\rm M} /\tau_{\rm M}$. Then, $I_s^{\rm back}$ can be written as
\begin{equation}
  I_s^{\rm back} = \frac{4 J_{\rm sd}^2 \chi_{\rm M}}{\hbar^2 \tau_{\rm M}} m_{\rm eq} \kB T_{\rm M} {\cal I}_1, \label{eq:I1def01} 
\end{equation}
where the integral ${\cal I}_1$ is defined by
\begin{equation}
  {\cal I}_1 = \int_\omega |g(\omega)|^2 {\rm Im}G_m(\omega). 
\end{equation}
We use the expression 
\begin{equation}
  {\rm Im}G_m(\omega) =
  -\frac{\gamma_m (\omega-d)(\omega-d^*)+ \gamma_n bc}{(\omega-\lambda_+)(\omega-\lambda^*_+)(\omega-\lambda_-)(\omega-\lambda^*_-)}, 
\end{equation}
where we introduced $\gamma_m = - {\rm Im}\, a= \Gamma_m/\widetilde{\chi}_{\rm AFI}$ and $\gamma_n = - {\rm Im}\,d= \Gamma_n K$. To proceed further, it is convenient to introduce the following notation:
\begin{eqnarray}
  \sqrt{(a-d)^2+ 4bc} &=& \sqrt{Z} = X + \ui Y, \\
  a+ d &=& \Omega- \ui \Gamma_+,
\end{eqnarray}
where $X$, $Y$, and $\Omega$ are pure real numbers, and
\begin{equation}
  \Gamma_\pm= \gamma_m \pm \gamma_n. 
\end{equation}
We evaluate the integral over $\omega$ by picking up the magnon poles $\omega= \lambda_\pm^*$, where we assume $\tau_{\rm M}^{-1} \gg \gamma_m, \gamma_n$ which means that the antiferromagnetic magnons are well-defined excitations. Then, after calculating the residues at $\omega= \lambda_\pm^*$, the integral is calculated to be
\begin{eqnarray}
  {\cal I}_1 &=& \frac{-\tau^2_{\rm M} }{\sqrt{Z^*}} \left[ \frac{p \lambda_+^*+ q}{(\Gamma_+-Y)(X+ \ui Y)} 
  - \frac{p \lambda_-^* + q}{(\Gamma_++Y)(X-\ui \Gamma_+)} \right] \nonumber \\
  &=&
  -\tau^2_{\rm M} \frac{\cal N}{\cal D}, 
\end{eqnarray}
where we defined
\begin{eqnarray}
  {\cal N} &=& p XY+ p \Gamma_+ (a^*+ d^*)+ 2 q \Gamma_+ , \\
  {\cal D} &=& (\Gamma^2_+- Y^2)(X^2 + \Gamma^2_+), \label{eq:calD01}
\end{eqnarray}
and we introduced $p= \gamma_m (a^*-d)$ and $q= -\gamma_m(a^*-d)d^*+ bc \Gamma_+$. Note that in the above equations, we used the relations $\lambda_+ + \lambda_- = a+d$ and $\lambda_+- \lambda_-= \sqrt{Z}$. The remaining task is to expand both ${\cal N}$ and ${\cal D}$ up to the second order with respect to $\gamma_m$ and $\gamma_n$. Using the relations $X^2- Y^2  \approx (a-d)^2+ 4bc$ and $XY \approx -(a-d)\Gamma_-$, the numerator and the denominator are calculated to be
\begin{eqnarray}
  {\cal N} &\approx& \frac{1}{2} \left\{ (\Gamma_+^2- \Gamma_-^2)(a-d)^2 + 4bc \Gamma_+^2 \right\}, \\
  {\cal D} &\approx& (\Gamma_+^2- \Gamma_-^2)(a-d)^2 + 4bc \Gamma_+^2, \label{eq:calD02}
\end{eqnarray}
where the higher-order corrections with respect to $\gamma_m, \gamma_n$ are disregarded. Substituting this result into Eq.~(\ref{eq:I1def01}), we obtain Eq.~(\ref{eq:IbackAFI02}).

Next, we come back to $I_s^{\rm pump}$ in Eq.~(\ref{eq:IpumpAFI01}). It is convenient to divide this quantity as
\begin{equation}
  I_s^{\rm pump} = I_{s,m}^{\rm pump}+ I_{s,n}^{\rm pump}, 
\end{equation}
where
\begin{eqnarray}
  I_{s,m}^{\rm pump} &=& \frac{-J_{\rm sd}^2 \chi_{\rm M}}{\hbar \tau_{\rm M}} \int_\omega
  |g(\omega)|^2 \omega |G_m(\omega)|^2  \bra \bra \xi^{-}_{\omega} \xi^{+}_{-\omega} \ket \ket,  \qquad \\
  I_{s,n}^{\rm pump} &=&  \frac{-J_{\rm sd}^2 \chi_{\rm M}}{\hbar \tau_{\rm M}} \int_\omega
  |g(\omega)|^2 \omega |G_n(\omega)|^2 
  \bra \bra \eta^{-}_{\omega} \eta^{+}_{-\omega} \ket \ket. \qquad 
\end{eqnarray}
We first calculate $I_{s,n}^{\rm pump}$ because it is easier to evaluate. Recalling the noise correlator in Eq.~(\ref{eq:etaeta}) with its spectral representation given by $\bra \bra \eta^{-}_{\omega} \eta^{+}_{-\omega} \ket \ket = 4 \kB T_{\rm AFI} \Gamma_n /\veps_0 v_0 $, $I_{s,n}^{\rm pump}$ can be written as 
\begin{equation}
  I_{s,n}^{\rm pump} = -\frac{4 J_{\rm sd}^2 \chi_{\rm M} \Gamma_n}{\hbar \veps_0 v_0} \kB T_{\rm AFI} {\cal I}_2, \label{eq:I2def01} 
\end{equation}
where the integral ${\cal I}_2$ is defined by
\begin{equation}
  {\cal I}_2 = \int_\omega |g(\omega)|^2 \omega |G_n(\omega)|^2, 
\end{equation}
with the last term of the integrand given by 
\begin{equation}
  |G_n(\omega)|^2 = 
  \frac{b^2}{(\omega-\lambda_+)(\omega-\lambda^*_+)(\omega-\lambda_-)(\omega-\lambda^*_-)}. 
\end{equation}
The integral over $\omega$ can be evaluated as before by picking up the magnon poles $\omega= \lambda_\pm^*$. After calculating the residues at $\omega= \lambda_\pm^*$, the integral is calculated to be
\begin{eqnarray}
  {\cal I}_2 &=& \frac{b^2 \tau^2_{\rm M}}{\sqrt{Z^*}} \left[ \frac{ \lambda_+^*}{(\Gamma_+-Y)(X+ \ui Y)} 
  + \frac{\lambda_-^*}{(\Gamma_++Y)(X-\ui \Gamma_+)} \right] \nonumber \\
  &=&
  \frac{ b^2 \tau^2_{\rm M} \left[ \Gamma_+ (a^*+ d^*)+ XY \right] }{ (\Gamma^2_+- Y^2)(X^2 + \Gamma^2_+) }. 
\end{eqnarray}

Calculation of $I_{s,m}^{\rm pump}$ can be done in a similar way. We recall the noise correlator in Eq.~(\ref{eq:xixi}) with its spectral representation given by $\bra \bra \xi^{-}_{\omega} \xi^{+}_{-\omega} \ket \ket = 4 \kB T_{\rm AFI} \Gamma_m /\veps_0 v_0 $. Then, $I_{s,n}^{\rm pump}$ can be written as 
\begin{equation}
  I_{s,m}^{\rm pump} = -\frac{4 J_{\rm sd}^2 \chi_{\rm M} \Gamma_m}{\hbar \veps_0 v_0} \kB T_{\rm AFI} {\cal I}_3, \label{eq:I3def01} 
\end{equation}
where the integral ${\cal I}_3$ is defined by
\begin{equation}
  {\cal I}_3 = \int_\omega |g(\omega)|^2 \omega |G_m(\omega)|^2, 
\end{equation}
with the last term of the integrand given by 
\begin{equation}
  |G_m(\omega)|^2 = 
  \frac{(\omega-d)(\omega-d^*)}{(\omega-\lambda_+)(\omega-\lambda^*_+)(\omega-\lambda_-)(\omega-\lambda^*_-)}. 
\end{equation}
The integral over $\omega$ can be evaluated as before by picking up the magnon poles $\omega= \lambda_\pm^*$, yielding 
\begin{eqnarray}
  {\cal I}_3 &=& \frac{\tau^2_{\rm M} }{\sqrt{Z^*}} \left[ \frac{k \lambda_+^*+ k}{(\Gamma_+-Y)(X+ \ui Y)} 
  + \frac{k \lambda_-^* + l}{(\Gamma_++Y)(X-\ui \Gamma_+)} \right] \nonumber \\
  &=&
  \frac{\tau^2_{\rm M} \left[ k XY+ k \Gamma_+ (a^*+ d^*)+ 2 l \Gamma_+ \right] }{ (\Gamma^2_+- Y^2)(X^2 + \Gamma^2_+) }, 
\end{eqnarray}
where we defined $k= a^*(a^*-d)+ bc$ and $l= (a^*-d)(-ad^*+ bc)$. 

Summing up $I_{s,m}^{\rm pump}$ and $I_{s,n}^{\rm pump}$, we obtain 
\begin{equation}
  I_s^{\rm pump}= -\frac{4 J_{\rm sd}^2 \chi_{\rm M} }{\hbar \tau_{\rm M} \veps_0 v_0} \kB T_{\rm AFI}
  \left( \Gamma_n {\cal I}_2+ \Gamma_m {\cal I}_3 \right).
  \label{eq:IsmIsn01}
\end{equation}
To proceed further, we first substitute $\Gamma_m= \gamma_m/r$ and $\Gamma_n= \gamma_n/K$, and we rewrite $\Gamma_n {\cal I}_2+ \Gamma_m {\cal I}_3$ as
\begin{eqnarray}
  \Gamma_n {\cal I}_2+ \Gamma_m {\cal I}_3  &=& \tau^2_{\rm M} \frac{\cal M}{\cal D}, \\
  {\cal M} &=& 
  \frac{\gamma_m}{r}\left\{ k XY+  \Gamma_+ [k(a^*+ d^*)+ 2 l ] \right\} \nonumber \\
  && + \frac{\gamma_n b^2 }{K} [ \Gamma_+ (a^*+ d^*)+ XY ], 
\end{eqnarray}
where ${\cal D}$ is defined in Eq.~(\ref{eq:calD01}). Then, we expand both ${\cal M}$ and ${\cal D}$ up to the second order with respect to $\gamma_m$ and $\gamma_n$. Using the relation $a \pm d \approx \gamma H_0 (1 \pm K/r)$ and $b^2/K= bc/r$, we obtain 
\begin{eqnarray}
  {\cal M} &\approx& \frac{\gamma H_0}{2 r} \left\{ (\Gamma_+^2- \Gamma_-^2)(a-d)^2 + 4bc \Gamma_+^2 \right\}, 
\end{eqnarray}
where ${\cal D}$ is given in Eq.~(\ref{eq:calD02}). Substituting the above result into Eq.~(\ref{eq:IsmIsn01}), we finally obtain Eq.~(\ref{eq:IpumpAFI02}), where Eqs.~(\ref{eq:m_eq02}) and (\ref{eq:chiAFI01}) are used.




\end{document}